# Surveying Students' Understanding of Quantum Mechanics

Chandralekha Singh and Guangtian Zhu

*Department of Physics and Astronomy, University of Pittsburgh, Pittsburgh, PA, 15260, USA*

**Abstract.** Development of conceptual multiple-choice tests related to a particular physics topic is important for designing research-based learning tools to reduce the difficulties. We explore the difficulties that the advanced undergraduate and graduate students have with non-relativistic quantum mechanics of one particle in one spatial dimension. We developed a research-based conceptual multiple-choice survey that targets these issues to obtain information about the common difficulties and administered it to more than a hundred students from seven different institutions. The issues targeted in the survey include the set of possible wavefunctions, bound and scattering states, quantum measurement, expectation values, the role of the Hamiltonian, time-dependence of wavefunction and time-dependence of expectation value. We find that the advanced undergraduate and graduate students have many common difficulties with these concepts and that research-based tutorials and peer-instruction tools can significantly reduce these difficulties. The survey can be administered to assess the effectiveness of various intructional strategies.

## INTRODUCTION

Learning quantum mechanics (QM) is challenging. The QM formalism is not very intuitive and it is very different from classical mechanics that students are used to from their everyday experiences and previous courses [1-6]. Moreover, a good understanding of the QM formalism requires a good grasp of mathematics including linear algebra, differential equations and special functions. Despite the mathematical facility required to master QM, the formalism of QM has a coherent conceptual framework. For learning to be meaningful, it is critical that the goals of a QM course, the instructional design and the assessment of learning are all aligned with each other. Since the students focus on what they are assessed on, the assessment of learning in QM should not only rely on measuring their facility with solving differential equations, it should also focus on their understanding of the conceptual framework and knowledge structure of QM. Without a coherent framework, students are unlikely to retain much after the QM course is over.

Research-based conceptual multiple choice surveys are useful tools for evaluating students' understanding of various topics [7]. The multiple choice surveys are easy to administer and grade. Their scores are objective and amenable to statistical analysis so that different instructional methods or different student populations can be readily compared. The Force Concept Inventory (FCI) is a conceptual multiple-choice test [8] that helped many instructors recognize that many introductory physics students were not developing a functional understanding of force concepts although they performed reasonably well on quantitative problems (often using an algorithmic approach). Other conceptual surveys have also been designed for many physics topics, e.g., electricity & magnetism [9]. These surveys reveal that students have many common conceptual difficulties with different topics in classical physics. Research-based instructional strategies have been shown to improve students' conceptual understanding of some of these topics significantly [10].

The conceptual difficulties that students have in the upper-level undergraduate courses (and even graduate students) are manifested even within the most stripped-down versions of quantum mechanics, i.e., non-relativistic quantum mechanics of one particle in one spatial dimension. We developed the Quantum Mechanics Survey (QMS) which is a 31-item multiple choice test covering various topics. The survey was developed by consulting with many QM instructors about the goals of their course, the topics their students should have definitely mastered and by iterating different versions of the open-ended and multiple-choice questions with a subset of them at various stages of the development of the survey. To investigate students' difficulties with various concepts, we administered free-response and multiple-choice questions and conducted interviews with individual students using a think-aloud protocol [11]. Individual interviews with the students during the investigation of the difficulties and the development of the survey were useful to obtain an in-depth understanding of students' thought processes.

## SURVEY DESIGN

The QMS focuses on assessing students' understanding of the conceptual framework of QM instead of assessing their mathematical skills. One can reason about all of the questions in the QMS conceptually and one need not necessarily perform any complicated integrals in order to answer them. Since the QMS focuses on quantum systems in one spatial dimension, the concept of orbital angular momentum was excluded from the survey. We also deliberately excluded the spin angular momentum and the Dirac notation from the QMS to ensure that the survey can be used after most of the first semester junior-senior level QM courses regardless of the textbook, the institution or the instructor.

We paid particular attention to the issues of reliability and validity [7] while designing the QMS. Reliability refers to the relative degree of consistency between the test scores if an individual repeats the test procedures. Validity refers to the appropriateness of interpreting the test scores. To ensure that the survey is valid, the opinions of 12 instructors about the goals of a junior-senior level QM course and the concepts their students should have definitely learned were taken into account during the development of the QMS. Apart from asking the instructors about these issues in online surveys, we discussed these issues individually with several instructors at the University of Pittsburgh (Pitt) who had taught QM at the junior-senior undergraduate level and/or at the graduate level.

The QMS includes a wide range of topics that the instructors expected their students to know such as the set of possible wavefunctions for a quantum system, the expectation value of a physical observable and its time dependence, the role of the Hamiltonian of a system, the stationary states and the non-stationary states and issues related to their time development, and quantum measurements. The quantum mechanical models in the QMS are all confined to one spatial dimension (1D), e.g., the 1D infinite/finite square well, the 1D simple harmonic oscillator (SHO) and the free particle. Before developing the questions for the QMS, we first developed a test blueprint based upon the instructors' feedback which provided a framework for deciding the desired test attributes. The specificity of the test plan helped us to determine the extent of content covered and the cognitive complexity levels of the questions.

In developing good alternative choices for the multiple-choice questions, we took advantage of the prior investigations of advanced students' difficulties with various topics at the junior/senior level QM, e.g., the set of possible wavefunctions, quantum measurements, time dependence of the wavefunction and expectation values [1-6]. The alternative choices for each question often had distractors which reflected students' common misconceptions to increase the discriminating properties of the questions. Having good distractors in the alternative choices is important so that the students do not select the correct answer for the wrong reason. To investigate students' difficulties further with some concepts before designing the multiple-choice questions for the QMS, we developed and administered to the students open-ended (or free-response) questions. The answers to the open-ended questions were summarized and categorized and helped develop good alternative choices for the multiple-choice questions in the QMS. Statistical analysis such as distribution of choices and correlation between distractors was conducted on the multiple-choice questions as they were developed and refined.

We also interviewed individual students using a think-aloud protocol [11] to develop a better understanding of students' reasoning process when they were answering the open-ended and multiple-choice questions. During these interviews, some previously unnoticed difficulties and misconceptions were revealed. These common difficulties were incorporated into the newer version of the written tests and ultimately into the multiple-choice questions in the survey developed. Four professors at Pitt reviewed the different versions of the QMS several times to examine its appropriateness and relevance for the upper-level undergraduate QM courses and to detect any possible ambiguity in item wording. Many professors from other universities have also provided valuable comments to fine tune the survey. Some of the questions were inspired by the learning tools for QM such as the concept tests and Quantum Interactive Learning Tutorials (QuILTs) we have developed [6,12]. Students' feedback to these questions is also an important resource for us to improve the clarity of QMS. Since we wanted the QMS to be administered within one class period, the final version of the QMS is limited to 31 multiple-choice questions. Each question has one correct and four alternative choices.

## THE SURVEY RESULTS

The QMS was administered to 109 students from seven universities (8 different classes were involved since both the upper-level undergraduate and graduate classes took the QMS at one institution). Among the 109 students, 15 were first-year graduate students enrolled in a full year graduate QM course. The others were undergraduate students who had taken at least a one-semester QM course at the junior-senior level. One of these junior-senior level classes in which students were enrolled for a full-year course used research-based learning tools such as concept tests and QuILTs [6,12]. The QMS was given twice to this class, at the end of the first semester (11 students) and at the end of the second semester (9 students).

The average score on the QMS for 109 students (only includes the first score of the students who took it twice) is 37.5%. The reliability coefficient [7] $\alpha$ for the survey is 0.87 which is reasonably good from the standards of test design. The item difficulty of each question (percentage of students who correctly answered each question), shown in Fig. 1, approximately ranged between 0.2 and 0.8. Most of the item difficulties (26 out of 31) were below 0.5. Fig. 2 shows the item discrimination which represents the ability of a question to distinguish between the high and low performing students in the overall test. One measure of item discrimination is the point biserial discrimination (PBD) coefficient [7], which is the correlation between "the score on a particular question" for each student and "the total test score minus the score on that question" for each student. The PBD approximately ranged from 0.2 to 0.6 with half of the questions with PBD higher than 0.4 and two items with PBD lower than 0.2. The standards of test design [7] indicate that the QMS questions have reasonably good PBD.

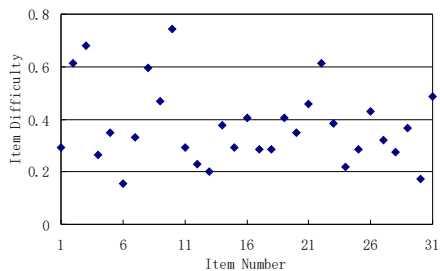

**Fig 1.** Item Difficulty (fraction correct) for each item on the test for 109 students

The average score for the upper-level undergraduate class that used the concept tests and the QuILTs throughout the semester was 71.8% at the end of the first semester in which all of the relevant concepts were covered and 74.9% at the end of the second semester of QM. During the second (spring) semester, these students were enrolled in the QM II course, which included identical particles and approximate methods such as the time-independent and dependent perturbation theories, variational method, scattering theory and WKB approximation. The course did not directly involve the contents in the QMS. It is encouraging that the average student performance did not deteriorate after a whole semester. In classes that did not use the research-based learning tools, the average score was 48.4% for the graduate course (15 students) and 31.0% for the undergraduate courses. The reliability coefficients $\alpha$ were greater than 0.8 for both the class that used the research-based learning tools and the group that did not use them.

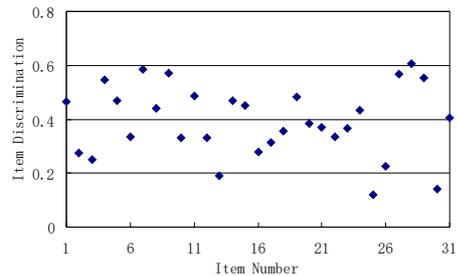

**Fig 2.** Item Discrimination for each item on the test for 109 students

## ITEM ANALYSIS

Table 1 shows one particular categorization of the questions in the QMS based upon the concepts. The table provides only one of the several possible ways to classify the questions. Our prior research shows [13] that different instructors categorize a given QM question in different ways so the categorization shown in Table 1 is only one of them that we found convenient. The group "Other" includes questions about the uncertainty principle, the concept of degeneracy in the context of a free particle, and the Ehrenfest theorem that says that the expectation value of a physical observable obeys the classical laws [1-6]. Below, we summarize the common difficulties found via the QMS in the sub-category of the time dependence of expectation values.

| Concepts | Number |
|---|---|
| Possible Wavefunctions | 5 |
| Bound/Scattering States | 5 |
| Measurement | 9 |
| Expectation Values | 3 |
| Time Dependence of expectation values | 4 |
| Stationary vs. Non-Stationary States | 8 |
| Role of the Hamiltonian | 3 |
| Time Dependence of Wavefunction | 7 |
| Other | 3 |

**Table 1.** One possible categorization of the QMS questions and the number of questions belonging to each category

Table 2 shows the percentages of students selecting the choices A-E on two problems related to the time dependence of expectation values. The correct responses are in bold italicized font. X refers to the percentage of students who did not attempt that question (left that question blank).

|  | A | B | C | D | E | X |
|---|---|---|---|---|---|---|
| Q2 | 8% | 7% | 5% | 17% | ***63%*** | 0% |
| Q23 | 8% | 9% | 14% | ***41%*** | 18% | 10% |

**Table 2.** Students' responses for the questions about two time dependence of expectation values

Questions (2) and (23) ask students about the time dependence of the expectation values of different physical observables in a stationary or a non-stationary state respectively. The questions are as follows:

Q2. *Suppose that at time t=0, System I (1D infinite square well) is in the first excited state. Choose all of the following expectation value(s) that depend on time.*
(1) <x>    (2) <p>    (3) <H>
A. 1 only    B. 2 only    C. 3 only    D. 1 and 2 only
E. None of the above

Q23. *Suppose at time t=0, System III (1D SHO) is in the state $(\psi_1 + \psi_2)/\sqrt{2}$. Choose all of the following expectation values that depend on time.*
(1) <x>    (2) <p>    (3) <H>
A. 1 only    B. 2 only    C. 3 only    D. 1 and 2 only
E. All of the above

In Question (2), the initial state is an energy eigenstate, so the expectation value of any time-independent operator is time-independent. The most common mistake in Question (2) was that the students believed that the expectation values of position and momentum depend on time in a stationary state. The initial state in Question (23) is a linear superposition of the stationary states $(\psi_1 + \psi_2)/\sqrt{2}$. The expectation value of energy is still time independent because the probability of obtaining energies $E_1$ or $E_2$ is always 50%. But the expectation values of position and momentum depend on time. Students need not evaluate the integrals to determine the correct response. Instead, if they realize that for a non-stationary state, the probability density changes with time, they can conclude that the expectation value of position and momentum must change with time. Another way to reason is to realize that the position and momentum operators do not commute with the Hamiltonian so their expectation values will depend on time in a non-stationary state. In Question (23), 18% of the students mistakenly thought that all the expectation values (position, momentum and energy) depend on time and 14% chose option C (only <H> depends on time) which is the opposite to the correct answer D (by contrast, only 5% of the students in Question (2) when the system was in a stationary state believed that <H> depends on time but <x> and <p> do not).

## SUMMARY

Identification of students' difficulties can help catalyze the design of better instruction strategies and learning tools to improve students' understanding of QM. We have developed a research-based multiple-choice survey to assess students' conceptual understanding of QM. The survey explores students' proficiency with the formalism of QM in 1D. During the development of the survey, we obtained feedback from QM instructors at various institutions, administered free-response and multiple-choice questions to students and conducted individual interviews with a subset of students to elucidate the difficulties students have with the concepts. The alternative choices for the multiple-choice questions on the survey often deal with the common difficulties found in these investigations.

The 31-item QMS was administered to 109 students in advanced undergraduate and graduate QM courses in seven different institutions to get a quantitative understanding of the universal nature of the difficulties. We found that the advanced students have common difficulties about various topics including the set of possible wavefunctions, quantum measurement, expectation values, stationary vs. non-stationary states, and time dependence of wavefunctions and expectation values. We also investigated the extent to which research-based learning tools can help students learn these concepts and found that the difficulties were significantly reduced when students used concept tests and QuILTs.

The QMS can be administered to students in the upper-level undergraduate courses after instruction. It can also be used as a preliminary test for the graduate students to evaluate their background knowledge in QM before they take the graduate-level QM courses. Those developing instructional strategies to improve students' understanding of QM can benefit from taking into account the difficulties highlighted by the QMS.

## ACKNOWLEDGMENTS

We thank the National Science Foundation for financial support and thank everybody who helped with the development and evaluation of the QMS.

## REFERENCES

1. C. Singh, Am. J. Phys. 69 (8), 885-896 (2001).
2. P. Jolly, D. Zollman, S. Rebello, and A. Dimitrova, Am. J. Phys. 66 (1), 57-63 (1998).
3. M. Wittmann, R. Steinberg, and E. Redish, Am. J. Phys. 70 (3), 218-226, (2002).
4. C. Singh, M. Belloni, and W. Christian, Physics Today 8 43-49, (2006).
5. C. Singh, Am. J. Phys. 76 (3), 277-287 (2008).
6. C. Singh, Am. J. Phys. 76 (4), 400-405 (2008).
7. G. Aubrecht and J. Aubrecht, Am. J. Phys. 51, 613-620 (1983); A. Nitko, Prentice Hall, Englewood, (1996).
8. D. Hestenes, M. Wells, and G. Swackhamer, Physics Teacher 30, 141-151 (1992)
9. D. Maloney, T. O'Kuma, C. Hieggelke, and A. Heuvelen, Am. J. Phys. 69, S12-S23 (2001).
10. R. Hake, Am. J. Phys. 66, 64 (1998).
11. M. T. H. Chi, Thinking Aloud, edited by Van Someren, Barnard, and Sandberg, Academic Press, NY, (1994).
12. E. Mazur, Peer Instruction: A User Manual, Prentice Hall, Upper Saddle River, NJ, (1997).
13. S. Y. Lin and C. Singh, Euro. J. Phys 31, 57-68 (2010).